# USING LINEAR PROGRAMMING TO CONSTRUCT BETTER CRITERIA FOR CLOSING THE DETECTION LOOPHOLE IN EPR EXPERIMENTS

James H. Bigelow
jlandjhb@earthlink.net
May 3, 2008

**ABSTRACT**

I formulate the problem of closing the detection loophole as a constrained optimization problem. Numerical methods can then be used to maximize the detector efficiency subject to the constraint that there exists a local realist explanation for the quantum correlations observed in the EPR experiment in question. Any detector efficiency larger than this maximum rules out all local realist explanations, and hence closes the detection loophole.

**INTRODUCTION AND BACKGROUND**

Quantum theory predicts that measurements performed on separated particles from an entangled system can have correlations that cannot be explained by any local realist (LR) model (I will shortly make this term precise). Experiments to test this prediction (called EPR experiments after the famous 1935 paper by Einstein, Podolsky, and Rosen) appear to agree with quantum theory predictions, but due to various loopholes they still admit LR explanations.

The detection loophole has been the subject of considerable analysis. In all EPR experiments to date, only a fraction of the trials yield particle detections by all observers (i.e., coincidences). Correlations calculated from only the coincidences defy local realist explanations. But local realist explanations may become possible if non-coincidences are included in the analysis, and if one does not assume that the trials that yield coincidences are a fair sample of all the trials. Analysis of this loophole is aimed at estimating the minimum fraction of particles that must be detected in order to rule out local realist explanations of the results without having to make this fair sampling assumption. This paper presents a new method for estimating this critical fraction, a method more reliable and more flexible than the methods currently in use.

**Local Realist Models**

An EPR experiment consists of distributing particles from an entangled system to $N \geq 2$ observers. Each observer selects one of $K \geq 2$ measurements to perform and obtains one of $Z \geq 2$ results.[1] I define a *setting* to be an $N$-tuple of measurements, one for each observer, and an *outcome* to be an $N$-tuple of results, one for each of the selected measurements. In each run of this experiment, one chooses one of the $K^N$ *settings*, generates a large number of entangled systems, and collects frequencies for each of the $Z^N$ possible *outcomes*. Alternatively, for each of the $K^N$ settings, one can use quantum theory to calculate the probabilities of each of the $Z^N$ possible outcomes. The result is a vector $q$ of dimension $m = (K^N) \times (Z^N)$, consisting of all of these frequencies (if obtained from an actual experiment), or probabilities (if calculated for a thought experiment).

Let $P$ be the set of $m$-dimensional vectors of probabilities for this EPR experiment that are achievable by a local-realistic model (abbreviated LR-achievable). Peres (1999) expresses the mainstream view of what it means for a vector of probabilities to be LR-achievable: that the components of the vector can be derived as the appropriate marginal probabilities of a joint probability distribution over all combinations of results from all the measurements considered in the EPR experiment. There are $n = Z^{N \times K}$ possible combinations of results. Thus there is an $m \times n$ matrix $A$, specific to EPR experiments with $N$ observers, $K$ measurements per observer, and $Z$ results per measurement, such that a vector of probabilities $p \in P$ if and only if $Ax = p$ for some $n$-dimensional vector $x \geq 0$.

The statement that the result $q$ of an EPR experiment has no local realist explanation, then, is just the statement that there exists no $n$-dimensional vector $x \geq 0$ that satisfies $Ax = q$. This statement can be expressed in the much more commonly encountered dual form, namely there exists an $m$-dimensional vector $y$ such that $y^T q > 0$ and $y^T p \leq 0$ for every $p \in P$. The inequality $y^T q > 0$ is, of course, a Bell inequality. The condition $y^T p \leq 0$ for every $p \in P$ is equivalent to $y^T A \leq 0$.

---

[1] It is easy to generalize to experimental setups where different observers can choose from different numbers of measurements, and different measurements can have different numbers of results.

The detection loophole works by increasing from $Z$ to $Z+1$ the number of results each measurement can yield, the additional result being a failure to detect. Now there are $\tilde{n} = (Z+1)^{N \times K}$ possible combinations of results. For example, in the archetypal EPR experiment (which has $N = K = Z = 2$), the number of possible combinations of results increases from $n = 2^{2 \times 2} = 16$ to $\tilde{n} = 3^{2 \times 2} = 81$.[2]

In principle, the experiment should now produce $(Z+1)^N$ joint frequencies from each of the $K^N$ settings, and the vector of probabilities $\tilde{q}$ should now have dimension $\tilde{m} = K^N \times (Z+1)^N$. Following the same procedure as before, define a new, larger $\tilde{m} \times \tilde{n}$ matrix $\tilde{A}$. In this expanded context, the vector $\tilde{q}$ is LR-achievable if there is a $\tilde{n}$-dimensional vector $\tilde{x} \geq 0$ that satisfies $\tilde{A}\tilde{x} = \tilde{q}$. It is not LR-achievable if there is no such $\tilde{x}$, or equivalently if $\tilde{q}$ satisfies a Bell inequality, i.e., there is a $\tilde{m}$-dimensional vector $\tilde{y}$ such that $\tilde{y}^T \tilde{q} > 0$ and $\tilde{y}^T \tilde{A} \leq 0$.

In practice, however, one discards trials of the experiment in which one or more observers get a "no detect" result (and in any case the outcome in which all observers get a "no detect" result is inherently unobservable). Nor can one calculate probabilities of non-detection from quantum theory. So where does the $\tilde{m}$-dimensional vector $\tilde{q}$ come from?

**Previous Models of the Detection Loophole**

Previous authors have obtained $\tilde{q}$ by modifying the vector $q$ of quantum probabilities in the case of perfect detection according to an ad hoc model of imperfections in the detection process, i.e., $\tilde{q} = \tilde{q}(q, u)$, where $u$ is a vector of parameters in the detection model. Strictly speaking, the detection loophole involves only false negatives, i.e., detectors failing to detect some of the particles sent to them. But the authors cited here often treat false positives (i.e., noise) in the same framework.

---

[2] Collins et al (2001) point out this approach to accommodating more than two results per measurement, but they do not take the step of considering a "no detect" to be one of the possible results.

Many authors introduce a single detection efficiency parameter, and assume that each detector registers a fixed fraction of the particles that encounter it, independently of whether or not other detectors register anything (e.g., Larsson & Semitecolos, 2000; Massar, 2001; Massar & Pironio, 2003). Some authors allow the detectors of different observers to have different efficiencies (Brunner et al, 2007; Cabello & Larsson, 2007). Massar et al (2002) include a parameter for the efficiency of producing particles at the source as well as the efficiency of detecting them. Durt, Kaszlikowski & Zukowski (2001) and Kaszlikowski et al (2000) include a noise parameter. Larsson (1999) and Zukowski et al (1999) introduce a visibility parameter, approximately equivalent to noise.

These authors then attempt to determine the most extreme values of $u$ for which $\tilde{q}$ is LR-achievable. If $u$ consists of only the single detection efficiency parameter, for example, the author estimates the largest detection efficiency that admits a local realist explanation of $\tilde{q}$. If investigators could achieve higher detection efficiencies, they conclude, the detection loophole would be closed.

These authors employ two approaches to estimate the local-realist limits of $u$. Using the first approach, Cabello, Rodriguez & Villanueva (2007) and Massar & Pironio (2003) construct solutions $\tilde{x} \geq 0$ to $\tilde{A}\tilde{x} = \tilde{q}(q,u)$ for particular $u$. This establishes a bound on the extremal $u$, and Cabello, Rodriguez & Villanueva (2007) (but not Massar & Pironio, 2003) are able to show that the bound is tight.

Authors using the second approach select a Bell inequality, i.e., a vector $\tilde{y}$ such that $\tilde{y}^T \tilde{q} \leq 0$ for all LR-achievable $\tilde{q}$, and solve for a value of $u$ at which the $\tilde{y}^T \tilde{q}(q,u) = 0$ (Brunner et al, 2007; Cabello & Larsson, 2007; Larsson, 1999; Larsson & Semitecolos, 2000; Massar et al, 2002). Of these authors, only Massar et al (2002) use Bell inequalities that include terms for "no detect" events. The others insert only the adjusted probabilities $\tilde{q}$ for which all observers get detections (i.e., coincidences) into traditional Bell inequalities. It is not immediately clear that this is wrong, but so far as I am aware, nobody has yet demonstrated that it is right. Note also that even if $\tilde{y}^T \tilde{q}(q,u) = 0$ for one $\tilde{y}$ corresponding to a Bell inequality, $\tilde{q}(q,u)$ need not be LR-achievable. In principle, it could violate another Bell inequality.

**NEW APPROACH TO MODELING EPR EXPERIMENTS WITH IMPERFECT DETECTION**

I suggest two improvements over the models and methods of the authors cited above. First, instead of developing special purpose, algebra-intensive methods to estimate critical values of $u$, one can use the powerful numerical methods of linear or non-linear programming. One would employ these methods to solve specific instances of the problem:

(1) $$\begin{aligned} &\text{Maximize} \quad f(u) \\ &\text{s.t.} \quad \tilde{A}\tilde{x} = \tilde{q}(q,u) \\ &\qquad \tilde{x} \geq 0 \end{aligned}$$

Here, $f(u)$ could be any function of the model parameters $u$. For example, Zukowski et al (1999) used linear programming to maximize visibility. As I will show, one can also maximize detection efficiency. If there are several parameters, some can be held fixed (e.g., Alice's detection efficiency set to 100 percent) and others maximized (e.g., Bob's detection efficiency).

Second, I suggest replacing the models $\tilde{q}(q,u)$ of previous authors with models that assume less about $\tilde{q}$. Each of their models has a very small number of parameters $u$, so that for fixed $q$ the function $\tilde{q}(q,u)$ is restricted to a small dimensional manifold in a large dimensional space. I agree that the models are plausible. But the real reason for limiting the number of parameters $u$ is to make it practical to solve for extremal values of $u$ algebraically. Numerical methods can deal with models that restrict $\tilde{q}$ less.

The advantage of assuming less about $\tilde{q}$ is that one can make more general statements about how good detectors must be to rule out local realist explanations. This is reminiscent of the Bell Theorem (Bell, 1964), which shows that under the assumption of perfect detection (a fair sampling assumption works as well), *all* local realistic models, not just a particular few, are incompatible with quantum theory.

The remainder of this paper formulates just such a more general model of the detection loophole. This model will consider detection errors of only one kind, namely the possibility that an observer fails to detect a particle. I defer consideration of the other kind of error, that a detection event occurs in the absence of a particle, to a subsequent paper.

**Preliminaries**

Let $Obs$ be the set of observers, with generic element $i$, as in $i \in Obs$. For the archetypal EPR experiment with two observers, $Obs = \{Alice, Bob\}$.

Let $Meas_i$ be the set of measurements available to observer $i$, with generic element $k$, as in $k \in Meas_i$. For the archetypal experiment with two measurements per observer, we might have $Meas_{Alice} = \{A1, A2\}$ and $Meas_{Bob} = \{B1, B2\}$.

Let $Res_{ik}$ be the set of results one can obtain from measurement $k \in Meas_i$, with generic element $r$, as in $r \in Res_{ik}$. For an experiment with dichotomous measurements plus a "no detect" possibility, every $Res_{ik}$ will be the same: $Res_{ik} = \{U, D, N\}$ (U for "Up," D for "Down," and N for "No detect").

Let $Det_{ik} \subset Res_{ik}$ be the subset of results for which there is a detection, i.e., $Det_{ik} = \{U, D\}$.

In terms of the earlier discussion, $Obs$ contains $N$ elements, every $Meas_i$ contains $K$ elements, every $Res_{ik}$ contains $Z+1$ elements, and every $Det_{ik}$ contains $Z$ elements.

**Categories of Trials**

In a local realist model, the system of particles generated at each trial of the experiment must carry a "set of instructions" or the equivalent telling it how to respond to every measurement, whether or not the measurement is actually performed. Thus each combination of one result for each possible measurement corresponds to a category of trials that are equivalent for the purposes of this experiment. The set of all categories for an EPR experiment with imperfect detection is:

$$\tilde{J} = \prod_{i \in Obs} \left( \prod_{k \in Meas_i} Res_{ik} \right)$$

I will later have occasion to use the set $J$ of categories for an EPR experiment with perfect detection:

$$J = \prod_{i \in Obs} \left( \prod_{k \in Meas_i} Det_{ik} \right)$$

For the archetypal EPR experiment (with $N = K = Z = 2$, and therefore $Z + 1 = 3$), the set $\tilde{J}$ contains $(Z+1)^{N \times K} = 81$ elements. For example, each row of Table 1 is an element of $\tilde{J}$. The elements with no entries "N" are also elements of the set $J$.

**Table 1: Sample of Elements of $\tilde{J}$ for the Archetypal EPR Experiment**

| A1 | A2 | B1 | B2 |
|----|----|----|----|
| U | U | U | U |
| U | U | U | D |
| U | U | U | N |
| U | U | D | U |
| U | U | D | D |
| U | U | D | N |
| U | U | N | U |
| U | U | N | D |
| U | U | N | N |
| … | … | … | … |
| N | N | N | D |
| N | N | N | N |

As stated earlier, every local realist model for a given EPR experiment can be represented as a probability distribution over these categories. Denote by $\tilde{x}_j$ the probability a trial will belong to category $j \in \tilde{J}$. Because these are probabilities they must satisfy:

(2) $$\sum_{j \in \tilde{J}} \tilde{x}_j = 1$$
$$\tilde{x}_j \geq 0$$

**Settings and Outcomes**

For each experimental trial, each observer will select one of his measurements to perform. As before, we use the term *setting* to refer to one such selection of one measurement per observer. The set of all settings is:

$$S = \prod_{i \in Obs} Meas_i$$

For the archetypal EPR experiment, the set $S$ contains $K^N = 4$ elements, as shown in Table 2.

**Table 2: Settings for Archetypal EPR Experiment**

| Alice's selection | Bob's selection |
|---|---|
| A1 | B1 |
| A1 | B2 |
| A2 | B1 |
| A2 | B2 |

Now suppose a trial is performed for setting $s = (s_1, s_2, ..., s_N)$, where $s_i \in Meas_i$ is the measurement selected by observer $i$. If every result could be recorded, including "no detect" results, one would obtain one of the $N$-tuples of results $r = (r_{s_i}, r_{s_2}, ..., r_{s_N})$ from the set:

$$R_s = \prod_{i \in Obs} Res_{is_i}$$

For the archetypal EPR experiment there will be nine possible outcomes for a trial performed for a given setting, say $s = (A1, B1)$. Each outcome will consist of one of the three possible results of measurement $A1$ paired with one of the three possible results of measurement $B1$, as shown in Table 3:

**Table 3: Possible Outcomes for an Arbitrary Setting for the Archetypal EPR Experiment**

| Alice's Result | Bob's Result |
|---|---|
| U | U |
| U | D |
| U | N |
| D | U |
| D | D |
| D | N |
| N | U |
| N | D |
| N | N |

**Relating Probabilities of Categories to Frequencies of All Outcomes**

Let $\tilde{q}_{sr}$ be the frequency with which outcome $r \in R_s$ occurs in a long sequence of trials at setting $s$. Each of these frequencies can be expressed as the sum of the appropriate subset of the probabilities $\tilde{x}_j$, as follows. An element $j \in \tilde{J}$ can be thought of as a vector with $\sum_{i \in Obs} Meas_i$ components, say $j = (j_{ik})$. For each setting $s = (s_1, s_2, ..., s_N)$, define a projection operator $P_s$ that selects the components of $j \in \tilde{J}$ that correspond to the measurements selected for that setting, i.e., $P_s(j) = (j_{1s_1}, j_{2s_2}, ..., j_{Ns_N})$. Each $P_s(j)$, then, is an element of the set of outcomes $R_s$. The probabilities $\tilde{x}_j$ for $j \in \tilde{J}$ that contribute to a frequency $\tilde{q}_{sr}$ are those for which $P_s(j) = r$. Thus the frequencies $\tilde{q}_{sr}$ can be calculated as:

(3) $\quad \tilde{q}_{sr} = \sum_{\{j | P_s(j) = r\}} \tilde{x}_j \quad s \in S, r \in R_s$

In matrix form, the set of equations (3) becomes $\tilde{A}\tilde{x} = \tilde{q}$, which appeared earlier. As noted in the section on previous models of the detection loophole, not all components of $\tilde{q}$ can be measured in experiments or calculated using quantum theory. If no observer detects a particle, for example, the investigator may not even realize there has been a trial. The next section will relate $\tilde{q}$ to the vector of observable or calculable frequencies $q$.

Before turning to this, however, it is useful to point out some features of the set of equations (3). First, equations (3), taken with equation (2) (the normalization condition for $\tilde{x}$) imply that the frequencies for a setting sum to 1. I omit the proof in the general case (which is trivial) in favor of illustrating it for the archetypal EPR experiment.

Table 4 lists the nine categories $j \in \tilde{J}$ whose probabilities $\tilde{x}_j$ sum to $\tilde{q}_{sr}$ for setting $s = (A1, B1)$ and outcome $r = (U, U)$. The frequency $\tilde{q}_{sr}$ for each of the eight other outcomes is the sum of a different set of nine probabilities $\tilde{x}_j$, i.e., no two outcomes share any $\tilde{x}_j$. Thus the sum $\sum_{r \in R_s} \tilde{q}_{sr}$ can be calculated as the sum of all 81 of the $\tilde{x}_j$, which by equation (2) equals 1.

**Table 4: Elements of $\tilde{J}$ That Contribute to the Frequency $\tilde{q}_{sr}$ for Setting $s=(A1,B1)$ and Outcome $r=(U,U)$ of the Archetypal EPR Experiment**

| A1 | A2 | B1 | B2 |
|----|----|----|----|
| U  | U  | U  | U  |
| U  | U  | U  | D  |
| U  | U  | U  | N  |
| U  | D  | U  | U  |
| U  | D  | U  | D  |
| U  | D  | U  | N  |
| U  | N  | U  | U  |
| U  | N  | U  | D  |
| U  | N  | U  | N  |

Equations (3) also imply that $\tilde{q}$ obeys a variety of no-signaling conditions. Again I eschew a general proof in favor of an illustration for the archetypal EPR experiment. In this example, the no-signaling condition states that the marginal probability that Alice obtains a result "U" given that she performs measurement A1 should not depend on whether Bob performs measurement B1 or B2. That is:

(4) $\quad \tilde{q}_{(A1,B1)(U,U)} + \tilde{q}_{(A1,B1)(U,D)} + \tilde{q}_{(A1,B1)(U,N)} = \tilde{q}_{(A1,B2)(U,U)} + \tilde{q}_{(A1,B2)(U,D)} + \tilde{q}_{(A1,B2)(U,N)}$

It is a simple (but tedious) matter to write out the six sets of nine $\tilde{x}_j$, each of which sums to one of the six frequencies in this equation. Inspection will show that the left-hand and right-hand sides of equation (4) are each the sum of the same 27 $\tilde{x}_j$.

For EPR experiments with more than two observers equations (3) imply multi-observer variations on the no-signaling conditions. For example, the probability that Alice obtains a result "U" given that she performs measurement A1 *and* Bob obtains a result "U" given that he performs measurement B1 should not depend on which measurement Charlie performs. This can be demonstrated by writing down an equation like (4) that expresses this condition, and observing that the left-hand side and right-hand side are sums of the same set of $\tilde{x}_j$.[3]

**Tallied Frequencies**

I picture the investigator performing many trials of the experiment at each setting, and tallying the joint frequencies of selected outcomes. In the typical experiment, the investigator discards all outcomes for which one or more observer fails to detect a particle (i.e., non-coincident outcomes). For a setting $s = (s_1, s_2, ..., s_N)$, where $s_i \in Meas_i$ is the measurement selected by observer $i$, this set of outcomes is:

$$D_s = \prod_{i \in Obs} Det_{is_i}$$

---

[3] While every LR-achievable $\tilde{q}$ obeys the normalization and no-signaling conditions, the converse is not true (e.g., see Tsirelson, 1993).

Adenier & Khrennikov (2006) and Peres (1995) point out that the investigator can, and argue that he should, also tally frequencies of outcomes in which some observers get a "no detect," so long as at least one observer gets a detection. I agree. Investigators generally use the Fair Sampling assumption to justify discarding these non-coincident outcomes, but I want to avoid making assumptions whenever possible. In principle, then, the investigator could tally frequencies of outcomes in any set $\tilde{D}_s$ that satisfies:

$$D_s \subseteq \tilde{D}_s \subseteq R_s - (N, N, ..., N)$$

While frequencies of non-coincidences can be collected in an experiment, however, quantum theory offers no means to estimate them theoretically. To model an entirely theoretical EPR experiment, or a published experiment for which only frequencies of coincidences are reported, one has no choice but to set $\tilde{D}_s = D_s$ for all settings $s$.

Denote by $q_{sd}$ the relative frequency with which outcome $d \in \tilde{D}_s$ occurs among trials that use setting $s$. By construction:

$$\sum_{d \in \tilde{D}_s} q_{sd} = 1$$

For the archetypal EPR experiment in which the investigator collects only the frequencies of coincidences, each set $\tilde{D}_s = D_s$ will contain $Z^N = 4$ elements, as shown in Table 5:

**Table 5: Tallied Outcomes for an Arbitrary Setting Archetypal EPR Experiment**

| Alice's Result | Bob's Result |
|---|---|
| U | U |
| U | D |
| D | U |
| D | D |

Because the tallied frequencies are normalized so that the frequencies for a setting sum to one, a tallied frequency $q_{sd}$ is not equal to the actual frequency $\tilde{q}_{sd}$ for the same setting and outcome. It is necessary to define a new variable for each setting, say $v_s$, that can be interpreted as the probability that a trial will contribute to one of the tallied frequencies, if the setting is $s$. Then the relation between actual and tallied frequencies is:

(5) $$\begin{aligned}\tilde{q}_{sd} &= q_{sd} \times v_s \quad s \in S, d \in \tilde{D}_s \\ v_s &\geq 0 \quad\quad\quad\quad s \in S\end{aligned}$$

Recall that the authors cited earlier calculate $\tilde{q}$ as a function of $q$ and some parameters $u$. All of their functions necessarily satisfy equations (4), but they all make additional assumptions about the form that imperfections in detections must take. Our model needs no further assumptions. In this it is reminiscent of Bell's Theorem (Bell, 1964), which demonstrates that if one assumes perfect detection (a fair sampling assumption works as well), all local realist models—not just one or two particular models—are incompatible with quantum theory.

**When Are Tallied Frequencies LR-Achievable with Imperfect Detection?**

I define a set of tallied frequencies $q$ to be *LR-achievable with imperfect detection* if there is a solution $(\tilde{x}, \tilde{q}, v)$ to equations (2), (3), and (5) for which $v_s > 0$ for every setting $s$. The solution $(\tilde{x}, \tilde{q}, v)$ is a local realist model that explains $q$. The condition $v_s > 0$ ensures that for every setting there will be some trials that yield outcomes $d \in \tilde{D}_s$, and therefore contribute to the frequencies tallied for that setting. Without this condition there can be settings for which the experiment yields no valid data.

It is easy to demonstrate that if $\tilde{D}_s = D_s$ for every setting $s$ (i.e., the only tallied frequencies are frequencies of coincidences), then every set of tallied frequencies is LR-achievable with imperfect detection. Denote by $jspec(s,d)$ the element of $\tilde{J}$ that yields outcome $d \in \tilde{D}_s$ for measurements selected in setting $s$, and a "no detect" result for every measurement not selected in setting $s$. Table 6 shows these elements of $\tilde{J}$ for one setting in the archetypal EPR experiment.

**Table 6: Elements of $\tilde{J}$ That Yield Valid Trials for Only Setting (A1,B1) in the Archetypal EPR Experiment**

| A1 | A2 | B1 | B2 |
|----|----|----|----|
| U  | N  | U  | N  |
| U  | N  | D  | N  |
| D  | N  | U  | N  |
| D  | N  | D  | N  |

Since there are $K^N$ settings, set $v_s = K^{-N}$ for every setting $s$. Set $\tilde{x}_{jspec(s,d)} = q_{sd} \times v_s$, and $\tilde{x}_j = 0$ if $j \neq jspec(s,d)$ for any pair $(s,d)$. This shows that $q$ is LR-achievable with imperfect detection.

As I pointed out earlier, $\tilde{q}$ must satisfy both the normalization and no-signaling conditions, but note that this is not required of the tallied frequencies $q$. In the above construction of a local realist explanation $(\tilde{x}, \tilde{q}, v)$ for $q$, I did not assume that $q$ satisfies the no-signaling conditions. Adenier & Khrennikov (2006) have reported experimental data that appear to violate no signaling conditions. By construction $q$ will satisfy the normalization conditions, and I used this fact in setting $v_s = K^{-N}$. But I can eliminate even this requirement by making the $v_s$ small enough.

However, if $\tilde{D}_s \neq D_s$ for some settings $s$ (i.e., frequencies are tallied for some non-coincidences), then the set of frequencies may not be LR-achievable with imperfect detection. This should be no surprise, as the literature cited earlier demonstrates that detection efficiencies above a certain threshold preclude all local realist explanations, and high detection efficiencies imply low frequencies of non-coincidences. It is also a good reason to retain frequencies of non-coincidences in experiments. Excluding them reduces the possibility that the experimental data alone, without additional ad hoc assumptions such as fair sampling, can demonstrate non-locality.

**Local Realist Models with Perfect Detection**

Any solution $(\tilde{x}, \tilde{q}, v)$ to equations (2), (3) and (5) represents a local realist model with imperfect detection. If every observer detects a particle at every trial, then it is also a local realist model with perfect detection. More formally, the probabilities $PDet_{ik}$ that an observer $i \in Obs$ will get a detection when he performs measurement $k \in Meas_i$ can be calculated as:

$$(6) \qquad PDet_{ik} = \sum_{\{j | j_{ik} \in Det_{ik}\}} \tilde{x}_j$$

Then the solution $(\tilde{x}, \tilde{q}, v)$ will have perfect detection if $PDet_{ik} = 1$ for every $i \in Obs$ and $k \in Meas_i$. Theorem 1 provides another way to recognize solutions with perfect detection.

**Theorem 1:** Let $(\tilde{x}, \tilde{q}, v)$ solve equations (2), (3) and (5) with $v > 0$. Then $\tilde{x}_j = 0$ for all $j \in \tilde{J} - J$ if and only if all the $PDet_{ik} = 1$.

**Proof:** By equation (6), each $PDet_{ik}$ is a sum of a subset of the $\tilde{x}_j$. By equation (2), $PDet_{ik} \leq 1$.

Assume that some $PDet_{ik} < 1$. Then there must be some $\tilde{x}_j > 0$ that does not appear in equation (5), and which therefore has $j_{ik} \notin Det_{ik}$. But this means $j \in \tilde{J} - J$.

Conversely, assume $\tilde{x}_j > 0$ for some $j \in \tilde{J} - J$. Then there must be some component of $j$, say $j_{ik}$ such that $j_{ik} \notin Det_{ik}$. Thus $\tilde{x}_j$ does not appear in the equation (6) that defines $PDet_{ik}$, so $PDet_{ik} < 1$. **QED.**

Theorem 1 simply says that if every observer detects something on every trial, regardless of the measurement he elects to perform, then all of the categories of trials $j \in \tilde{J}$ that have a "no detect" result for any measurement must have probability zero. The converse is also true.

**Theorem 2:** Let $(\tilde{x}, \tilde{q}, v)$ solve equations (2), (3) and (5) with $v > 0$, and suppose that $\tilde{x}_j = 0$ for all $j \in \tilde{J} - J$. Then for every setting $s$:

1. $\tilde{q}_{sr} = 0$ for all $r \in R_s - D_s$.
2. $q_{sd} = 0$ for all $d \in \tilde{D}_s - D_s$ (this is trivially true if $\tilde{D}_s = D_s$).
3. $v_s = 1$.
4. $\tilde{q}_{sd} = q_{sd}$ for all $d \in D_s$.

**Proof:** Suppose the converse of statement 1. Then there is a setting $s$ and an outcome $r \in R_s - D_s$ for which $\tilde{q}_{sr} > 0$. There must therefore be some $j$ for which $P_s(j) = d$ and $\tilde{x}_j > 0$. But if $P_s(j) \in R_s - D_s$, then $j \in \tilde{J} - J$, contradicting the premise of the lemma. Thus statement 1 is true.

Next, statement 2. For every $d \in \tilde{D}_s$, we have $\tilde{q}_{sd} = q_{sd} \times v_s$. By statement 1, $\tilde{q}_{sd} = 0$ for every $d \in R_s - D_s \supset \tilde{D}_s - D_s$. Since $v_s > 0$, if $\tilde{q}_{sd} = 0$ then $q_{sd} = 0$.

To demonstrate statement 3, sum equations (5) for a setting $s$ over all outcomes $d \in D_s$. By statements 1 and 2, this sum includes all the positive $q_{sd}$ and $\tilde{q}_{sd}$ for this setting, and hence $\sum_{d \in D_s} q_{sd} = \sum_{d \in D_s} \tilde{q}_{sd} = 1$. The sum of equations (5) for this setting then simplifies to $v_s = 1$.

Statement 4 follows immediately from statement 3 and equations (6).
**QED.**

It follows, then, that if there is a local realist explanation with perfect detection for the tallied frequencies $q$, then those frequencies satisfy the no-signaling conditions discussed earlier. Bell's theorem (Bell, 1964), of course, shows that the converse is not true. There are probabilities $q$ that satisfy the no-signaling conditions but do not admit local realist explanations with perfect detection.

As mentioned earlier, if one does not assume perfect detection, then the frequencies $q$ can admit a local realist explanation $(\tilde{x}, \tilde{q}, v)$ even if they don't satisfy no signaling conditions and their generalizations to subsets containing multiple observers.

**EXAMPLES**

**Criteria for Ruling Out Local Realist Models with Imperfect Detection**

The primary usefulness of constructing a local realist model with imperfect detection is to learn how near to perfect detection must be to rule out all local realist explanations. To operationalize this notion, consider any function $f(\tilde{x}, \tilde{q}, v)$. By maximizing $f(\tilde{x}, \tilde{q}, v)$ subject to constraints (2), (3) and (5), one can find $f^*$ such that $f(\tilde{x}, \tilde{q}, v) \leq f^*$ for every local realist model $(\tilde{x}, \tilde{q}, v)$ with imperfect detection. If experimental evidence that shows $f(\tilde{x}, \tilde{q}, v) > f^*$ in the real world, the detection loophole for this EPR experiment will be closed.

There are many functions $f(\tilde{x}, \tilde{q}, v)$ one might choose to maximize. For my examples I choose functions of the detection probabilities $PDet_{ik}$ (equations (6)) that I can compare to some of the parameters other authors have used in their models of imperfect detection. In addition, these functions permit me the convenience of using linear programming[4] to find $f^*$.

Define a new variable for each observer, $dmin_i$, equal to the minimum probability of detection over all measurements that observer might perform. This requires the following additional constraints:

(7)     $dmin_i \leq PDet_{ik} \quad i \in Obs, k \in Meas_i$

It may appear that $dmin_i$ can actually take on a value strictly less than the minimum of the $PDet_{ik}$, but maximizing $dmin_i$ (or any function that is increasing in $dmin_i$) will force $dmin_i$ to equal the minimum. That is, it is *feasible* for $dmin_i$ to be strictly smaller than the minimum, but it is not optimal.

**Examples of EPR Experiments with Two Observers**

For our two-observer examples, I examine three criteria for ruling out local realist explanations. First I maximize the smaller of $dmin_{Alice}$ and $dmin_{Bob}$. This requires adding a new variable, $dsym$, and some additional constraints:

(8)     $dsym \leq dmin_i \quad i \in Obs$

Then I maximize $dsym$ subject to (2), (3), (5), (6), (7) and (8) as constraints. This can be compared to other authors' results for assumptions about imperfect detection that are symmetric among the observers.

Second, I first maximize $dmin_{Alice}$ subject to (2), (3), (5), (6) and (7) as constraints. Suppose its maximum value is $dmin^*_{Alice}$ (as it happens, $dmin^*_{Alice} = 1$ in all of my examples). Then I impose one more constraint:

---

[4] Linear programming is convenient because it is a mature, well-understood methodology for solving such problems, e.g., see Hillier & Lieberman, 2005. Reliable software exists for solving very large linear programs, e.g., see GAMS, undated.

(9) $dmin_{Alice} \geq dmin_{Alice}^*$

and maximize $dmin_{Bob}$ subject to (2), (3), (5), (6), (7) and (9) as constraints. This can be compared to assumptions about imperfect detection that are asymmetric among the observers, such as Brunner et al (2007) and Cabello & Larsson (2007) considered.

Finally, I reverse the roles of Alice and Bob. That is, I maximize $dmin_{Bob}$, and subsequently maximize $dmin_{Alice}$ with $dmin_{Bob}$ constrained to its maximum value.

**Results for Experiments with $N = K = Z = 2$**

Table 7 shows results for four EPR experiments with two observers, two measurements per observer, and two results per measurement. The tallied frequency data for each example can be found in Appendix III of van Dam, Gill & Grunwald (2004). For example, in the "Optimized Bell" experiment for spin ½ particles, the four possible measurements are $(A1, A2, B1, B2) = \left(0, \frac{\pi}{3}, 0, \frac{2\pi}{3}\right)$ (half these angles for an experiment using photons). Only if both detectors register detections for more than 90 percent of trials will all local realist models be ruled out. But if Alice's detector registers detections for 100 percent of the trials, then all local realist models will ruled out if Bob's detector registers detection for more than 80 percent of trials.

**Table 7: Results for Four Experiments with $N = K = Z = 2$**

| Objective (to be maximized) | Original Bell | Optimized Bell | CHSH | Hardy |
|---|---|---|---|---|
| $dsym = MIN(dmin_{Alice}, dmin_{Bob})$ | 0.9142 | 0.9 | 0.8536 | 0.9236 |
| $dmin_{Alice}$ given $dmin_{Bob} = 1$ | 0.8284 | 0.8 | 0.7071 | 0.8472 |
| $dmin_{Bob}$ given $dmin_{Alice} = 1$ | 0.8284 | 0.8 | 0.7071 | 0.8472 |

The detection efficiencies obtained here are not identical to efficiencies reported in the literature, though they are not markedly different. For example, for experiments where Alice and Bob each choose between two measurements, Brunner et al (2007) and Cabello & Larsson (2007) estimate that a detection efficiency of at least 0.8284 is required to close the detection loophole for a symmetric experiment, and 0.7071 for an asymmetric experiment. For experiments where each observer chooses among three measurements, Massar et al (2002) estimate critical efficiencies of 0.8165 and 0.8217 for a symmetric experiment, while Brunner et al (2007) estimate a critical efficiency of 0.6667 for an asymmetric experiment.

Our results differ from one case to the next, showing that the critical efficiency to close the detection loophole depends on just what experiment is performed. This is true as well for estimates of critical efficiencies from the literature.

**Results for an Experiment with** $N=2, K=3, Z=2$

Zukowski et al (1999) suggest that allowing each observer to choose among all possible measurements—certainly more than two—would make for more general results. They seek to estimate the maximum visibility consistent with a local realist explanation, but the same suggestion can be applied to estimating critical detector efficiencies.

Table 8 shows the results for an experiment with two observers, each choosing among three measurements with two possible results, not counting "no detect" results. The tallied frequency data can be found in Appendix III of van Dam, Gill & Grunwald (2004), where this is labeled the Mermin experiment. In this experiment, the critical detection efficiencies are lower than for any of the two-measurement experiments.

**Table 8: Results for an Experiment with** $N=2, K=3, Z=2$

| Objective (to be maximized) | Mermin |
|---|---|
| $dsym = MIN(dmin_{Alice}, dmin_{Bob})$ | 0.8333 |
| $dmin_{Alice}$ given $dmin_{Bob} = 1$ | 0.6667 |
| $dmin_{Bob}$ given $dmin_{Alice} = 1$ | 0.6667 |

I generated about a dozen experiments at random with four measurements per observer, and an equal number with five measurements per observer. (The size of the linear program increases exponentially with the number of measurements per observer. I did not attempt more than five.) None of these cases improved on the results for the Mermin case. Obtaining lower critical detection efficiencies, then, requires picking the right measurements, not merely more measurements.

Incidentally, in some of the random cases, the asymmetric critical detection efficiencies were different for Alice than for Bob.

**Results for an Experiment with** $N = K = 2, Z = 3$

Kaszlikowski et al (2000) and Durt et al (2001) have shown that experiments with measurements with $Z \geq 3$ results violate local realism more strongly than experiments with two-valued measurements, and on this basis one might speculate that the detection loophole for such experiments could be closed at lower detection efficiencies. Their measure of the strength of LR violation is the maximal fraction of noise that one can mix with the quantum-theoretic frequencies and still rule out local realist explanation. I did a similar exercise using the maximal detector efficiency as the measure of the strength of the violation. The tallied frequencies $q$ for this example come from Kaszlikowski et al (2001). Table 9 shows the results. The results do not improve on Mermin, but they are superior to any of the examples in Table 7 (i.e., the $N = K = Z = 2$ examples).

**Table 9: The Qutrit Experiment**

| Objective (to be maximized) | Qutrit |
|---|---|
| $dsym = MIN(dmin_{Alice}, dmin_{Bob})$ | 0.8481 |
| $dmin_{Alice}$ given $dmin_{Bob} = 1$ | 0.6962 |
| $dmin_{Bob}$ given $dmin_{Alice} = 1$ | 0.6962 |

**Results for an Experiment with** $N = 3, K = Z = 2$

If increasing the number of measurements per observer does not guarantee better results, perhaps increasing the number of observers will. Table 10 shows the results for the Greenberger-Horne-Zeilinger (GHZ) experiment, the only experiment with more than two observers that van Dam, Gill & Grunwald (2004) considered. Instead of three objective functions there are seven:

- Maximize the minimum of the three observers' efficiencies (the symmetric objective);

- Set one observer's detection efficiency to 100 percent and maximize the minimum of the other two efficiencies (three objectives);
- Set two observers' efficiencies to 100 percent and maximize the efficiency of the remaining observer (three objectives).

**Table 10: Results for the GHZ Three-Observer Experiment**

| Objective (to be maximized) | GHZ |
|---|---|
| $dsym = MIN(dmin_{Alice}, dmin_{Bob}, dmin_{Charlie})$ | 0.8333 |
| $MIN(dmin_{Alice}, dmin_{Bob})$ given $dmin_{Charlie} = 1$ | 0.75 |
| $MIN(dmin_{Alice}, dmin_{Charlie})$ given $dmin_{Bob} = 1$ | 0.75 |
| $MIN(dmin_{Bob}, dmin_{Charlie})$ given $dmin_{Alice} = 1$ | 0.75 |
| $dmin_{Alice}$ given $dmin_{Bob} = dmin_{Charlie} = 1$ | 0.5 |
| $dmin_{Bob}$ given $dmin_{Alice} = dmin_{Charlie} = 1$ | 0.5 |
| $dmin_{Charlie}$ given $dmin_{Alice} = dmin_{Bob} = 1$ | 0.5 |

The results do not improve upon the two-observer Mermin experiment, unless one can devise an asymmetric experiment in which two of the observers have perfect detectors.

**SUMMARY AND NEXT STEPS**

I have described a method for developing criteria for closing the detection loophole in EPR experiments. According to the mainstream view, a local realist model of an EPR experiment is a probability distribution over all combinations of results of all measurements considered in the experiment. This distribution explains the EPR experiment if its marginal probabilities reproduce the frequencies of outcomes observed in the experiment. To address the detection loophole, I add a "no detect" result for each measurement. Finding a local realist model that explains an EPR experiment then becomes a matter of finding a solution in non-negative variables for a set of linear equations.

Logically, there can be two kinds of detection errors, false negatives and false positives. In this paper I have considered only false negatives, i.e., a failure to detect a particle that has arrived at the detector. This limitation has enabled me to formulate the problem as a linear program, and hence to use a widely available and reliable method for developing criteria for closing the detection loophole.

I close the paper with three suggestions for extending the method.

**Modeling False Positives**

Previous authors have considered false positive detections under the labels of noise (Durt, Kaszlikowski & Zukowski, 2001; Kaszlikowski et al, 2000) or visibility (Larsson, 1999; Zukowski et al, 1999). In each of these articles, the quantum probability vector $q$ is replaced with $V \times q + (1-V)$, where $V$ is the visibility (in the latter two articles) and $(1-V)$ is the noise parameter (in the former two articles). Thus (1) false positive detections are assumed to be statistically independent of true outcome, and (2) non-coincidences (some detectors firing while others remain silent) are not considered.

There should be little difficulty applying the method developed in this paper to this model of false positives. But false positives should sometimes occur at one detector but not others, producing outcomes $d \in \tilde{D}_s - D_s$. Incorporating information about the frequencies of these non-coincident outcomes will impose extra constraints on the local realist probability distribution $\tilde{x}$. It will be interesting to see how much this mutes the effect of false positives on the size of the detection loophole.

**Application to Experimental Data**

Experiments can produce frequencies of both coincident and non-coincident outcomes. Applying this method to experimental data, including frequencies of non-coincidences, might show that the detection loophole is smaller than current methods suggest.

**Design of EPR Experiments Resistant to Imperfect Detection**

The method described in this paper takes as inputs the probabilities $q$, and estimates how perfect detection must be to rule out all local realist explanations of $q$. But one might consider $q$, not as a constant vector, but as a function of the design of the EPR experiment. The design parameters would be additional variables in the method. One might, for example, seek values of those design parameters for which the maximum value of *dsym* is as small as possible.

**ACKNOWLEDGEMENTS**

I would like to thank Adan Cabello for very kindly endorsing me to submit papers to arXiv/quant-ph.